\documentstyle[preprint,aps,prl,epsfig]{revtex}

\tighten
\begin{document}
\draft
\preprint{version 1.0}

\title{ Complete Result for Positronium Energy Levels 
        at Order \boldmath{$\alpha^6 \, m$}}

\author{Krzysztof Pachucki$^{a b}$ 
\thanks{E-mail address: krp@fuw.edu.pl} and 
Savely G. Karshenboim$^{a c}$ 
\thanks{E-mail address: sgk@onti.vniim.spb.su}} 
\address{
$^a$ Max-Planck-Institut f\"{u}r Quantenoptik,
Hans-Kopfermann-Str. 1, 85748 Garching, Germany 
\thanks{temporary address},\\
$^b$ Institute of Theoretical Physics, Warsaw University,
Ho\.{z}a 69, 00-681 Warsaw, Poland,\\
$^c$ D.I. Mendeleev Institute for Metrology, Moskovsky pr. 19,
St. Petersburg 198005, Russia.}
\maketitle

\begin{abstract}
We have completed theoretical predictions for positronium energy levels 
through order $\alpha^6 \, m$ by the calculation 
of the spin independent, radiative recoil correction. 
This contribution is significant and amounts to $10.64$ MHz 
for the 1S state. We further
perform detailed comparison of theoretical predictions
to experimental results for $1S-2S$ and $2S-2P$ transitions. 
\end{abstract}
\narrowtext
Positronium is an unique hydrogenic atom to study quantum electrodynamic (QED)
effects in bound systems. Small mass of point--like leptonic constituents: 
electron and positron, causes that the strong and weak interaction effects
contribute at the negligible precision level. The energy spectrum
may therefore be predicted with the accuracy limited only
by the complexity in the higher order QED calculations. 
The equal mass of the electron and the positron requires
a special theoretical treatments to incorporate relativistic and recoil
effects on the same foot.  Moreover, the annihilation of positronium
affects also its energy levels. 
Therefore, the calculations of higher order corrections 
presents a challenge in the development of the quantum electrodynamics.
Recently a significant progress has been achieved
by the complete calculation of a single photon annihilation
contribution to the positronium hyperfine splitting ({\em hfs}) by
Adkins {\em et al.} \cite{Adkins971} and Hoang {\em et al.} \cite{Hoang},
and a pure recoil contribution \cite{Pachucki972} to S-levels energies.
In this letter we report on the calculation of the last unknown
correction to positronium S-levels in the order of $\alpha^6\,m$.
It is a spin independent radiative recoil contribution.
Having it evaluated, we give theoretical predictions in Table \ref{t2}, 
for six transitions in positronium
which are experimentally known with a high accuracy. In conclusions
we emphasize that current predictions are now more accurate than 
experimental values and in some cases a disagreement is met.

The radiative recoil corrections are in general difficult in treatment.
A rigorous derivation starts from the Bethe-Salpeter equation,
and incorporates radiative corrections in the kernel \cite{adkinsrev}.
In the order of $\alpha\,(Z\,\alpha)^5\,m$ 
a simplified treatment is sufficient. 
One finds an effective interaction potential between
the electron and the positron from the corresponding S-matrix amplitude.
When this amplitude is infrared divergent, than the separate
treatment is necessary for small photon momenta.
An illustrative example is the one--loop contribution
to the hydrogen Lamb shift of order $\alpha\,(Z\,\alpha)^4\,m$, 
where the low energy part leads to the well known Bethe logarithm.
In a higher order $\alpha\,(Z\,\alpha)^5\,m$ 
this low energy part is simply absent.
In the case of positronium we calculate the forward two-photon
exchange electron-positron scattering amplitude
at zero spatial components of the momenta. The nonperturbative effects 
enter only in the lower order. 
This simplified treatment has been justified in  Ref. \cite{Eides89}.
Similar calculations have  also been performed for
the positronium {\em hfs} in Ref. \cite{Terray},
and for radiative recoil corrections to hydrogen Lamb shift 
in  Ref. \cite{Pachucki951}.

We start the calculation from the vacuum polarization effect.
It  modifies one of the photon propagators, what gives a small energy shift. 
Our result is \cite{conventions}:
\begin{equation}
\Delta E(nS)=\frac{\alpha^6\, m}{n^3}\,
\left(\frac{1}{36}-\frac{5}{27\,\pi^2}\right)\approx\frac{\alpha^6\, m}{n^3}\,
0.00901
\,, \label{1}
\end{equation}
in agreement with the former result in  Ref. \cite{Eides}.
The electron and the positron self--energy correction 
to positronium S--levels could be written in the form \cite{elsewhere}
\begin{eqnarray}
\Delta E(nS) & = &  -\frac{\alpha^6\,m^3}{\pi\,n^3}\,
\int\frac{d^4q}{\pi^2\,i} \;{\cal L}(q)\,, \label{2}
\\ {\cal L}(q)  & = & 
\frac{q^2\,T^{0}_0(q) + 
q_0^{\,2}\,T^\mu_\mu(q)}{q^4\,(q^4-4\,m^2\,q_0^{\,2})}\,, 
\label{3}
\end{eqnarray}
where $T^{\mu}_{\nu}$ is a self energy correction to one-particle amplitude, 
and it was calculated analytically in Ref. \cite{Pachucki951}. The integration
is done along the Feynman contour. 
The expression in Eq. (\ref{3}) requires subtraction of terms, that were
included in the leading order self--energy contribution of order $\alpha^5\,m$,
\begin{eqnarray}
{\cal L} \approx
\frac{-\frac{10}{9}+\frac{4}{3}\,\ln\left[({\bbox q}^2-2\,m\,q_0)/m^2\right]}
{\pi\,({\bbox q}^2-2\,m\,q_0)^2\,({\bbox q}^2+2\,m\,q_0)}\,.
 \label{4}
\end{eqnarray}
The integral in Eq. (\ref{2}) with the subtracted ${\cal L}$ is finite
and amounts to 
\begin{equation}
\Delta E(nS)  = \frac{\alpha^6 \, m}{n^3}\,0.5612(1)\,. \label{5}
\end{equation}
All other corrections up to the order $\alpha^6 \, m$ have already been
calculated, so we could present an improved theoretical predictions.

The main structure of  the positronium spectrum 
is obtained from the nonrelativistic Hamiltonian. Leading relativistic
effects of order $\alpha^4 \, m$ are given for arbitrary positronium states 
by Ferrel formula.
Corrections of order $\alpha^5$ appear including Lamb shift-like effects,
recoil and annihilation contributions \cite{Karplus,Fulton}. 
Final results for them can be found for example in Ref. \cite{Fell}. 
Recently Khriplovich and coworkers have calculated all $\alpha^6 \, m$
corrections to P-levels in Ref. \cite{Khriplovich932} using the
Breit formalism.  In the case of $S$-states the calculation
is much more complex. The logarithmic terms 
$\alpha^6\,\ln(\alpha)\,m$ come from the 
single photon annihilation channel 
\cite{Caswell78}, 
and from spin dependent part of photon exchange contributions 
\cite{Lepage}. The spin independent part, as it was found in Refs. 
\cite{Fell,Khriplovich931}, does not lead to $\ln(\alpha)$ terms.
Values for nonlogarithmic corrections are displayed in Table \ref{t1}. We
divide them into 1--, 2--, 3--
photon annihilation terms, and 0--, 1--, 2-- radiative loop exchange terms.
Additionally, photon exchange terms have spin independent and
spin dependent part, proportional to $\bbox{s}_-\bbox{s}_+$ operator,
that leads to the hyperfine splitting.

The photon annihilation terms have been calculated for the {\em hfs}. 
The contribution to energy levels could be found, by noting that 
1-- and 3--photon annihilation contribute for the orthopositronium
and 2--photon for the parapositronium only, see Table \ref{t1}. 
The one-photon annihilation has been investigated in detail in
Refs. \cite{Barbieri,Samuel,Karshenboim932} 
and the calculation has been completed recently.
Most of the annihilation terms are state independent, 
{\em i.e.} they have only $1/n^3$ prefactor.
The one--photon annihilation brings a nontrivial state dependence.
The complete formula as derived by Hoang {\em et. al} in Ref. \cite{Hoang} is
\begin{eqnarray} \label{1-annih}
\Delta E(n^3S_1) = \frac{\alpha^6\, m}{n^3} 
&\biggl\{& \frac{1}{24}\ln(\alpha^{-1})\,-0.1256487 
\nonumber \\ & & 
+\frac{1}{24}\bigl[
\ln(n) -\Psi(n)+\Psi(1)\bigr] \nonumber \\ & & 
+\frac{1-n}{24\,n}-\frac{37}{96}\,\frac{1-n^2}{n^2}
\biggr\}\, ,
\end{eqnarray}
where $\Psi(n)$ is the logarithmic derivation of the Euler $\Gamma$-function.
The value for $n=1$ has been independly calculated by Adkins {\em et al.}
\cite{Adkins971},
what forms a crucial check for this complicated calculation.
Since the state dependent part is surprisingly large 
we recalculated it and got an agreement with Eq. (\ref{1-annih}).
The state dependence of the vacuum polarization contribution 
was calculated also in Ref. \cite{Abalmasov}.
Two-photon and three-photon annihilation contributions are state 
independent and the final results after correcting previous calculations
are presented  in Refs. \cite{Adkins88} and \cite{Adkins93} respectively.

The photon exchange contributions to the hyperfine structure
is presented as a spin dependent part in Table \ref{t1}.
Two--loop radiative term
is given by $\alpha^2$ part of $(1+a_e)^2$,
where $a_e$ is the electron anomalous magnetic moment.
Single radiative loop exchange, {\em i.e.} radiative recoil 
correction ($\alpha\,(Z\alpha)^5\,m$) was found in Ref.
\cite{Terray}. 
We have checked this result using the method of Ref. \cite{Pachucki951}
and obtained agreement.
Pure recoil contribution ($(Z\alpha)^6\,m$),
which is state dependent,  
was evaluated in Ref. \cite{Caswell86} for $n=1$, 
but recently it has been recalculated for all $nS$ states
in Ref.  \cite{Pachucki971}
\begin{eqnarray}
\Delta E(nS)  =  \frac{\alpha^6\,m}{n^3}\,\bbox{s}_-\bbox{s}_+\,
&\biggl\{& \frac{1}{6}\,\ln(\alpha^{-1})\,+0.3767(17)
\nonumber \\ & & 
+\frac{1}{6}\bigl[\ln(n)-\Psi(n)+\Psi(1)\bigr]
\nonumber \\ & & 
+\frac{7}{12}\,\frac{1-n}{n}-
\frac{1-n^2}{2\,n^2}
\biggr\}\, , 
\end{eqnarray}
where
\begin{eqnarray}
\bbox{s}_-\bbox{s}_+ = \left\{
\begin{array}{r}
1/4 \;\mbox{\rm for triplet,}\\
-3/4 \;\mbox{\rm for singlet.}
\end{array}
\right. \label{8}
\end{eqnarray}
These two results [0.167(33) versus 0.3767(17)] are in straight 
disagreement at $n=1$.
In Table \ref{t1} and below, we include results of Ref. \cite{Pachucki971}, 
but in Table \ref{t2} both contradicting results 
are presented for the ground state {\em hfs}. 

The spin independent part of the photon exchange contributions  
can be also classified as
$\alpha^2(Z\alpha)^4\,m$, $\alpha(Z\alpha)^5\,m$ and $(Z\alpha)^6\,m$
terms. The first one, a single photon exchange,  
can be found from the well-known two-loop expression 
for the hydrogen Lamb shift, taking into account the
reduced mass effects and multiplying the self-energy by factor 2.
The radiative recoil contribution  $\alpha(Z\alpha)^5m$ is calculated 
in this work. It is a sum of Eqs. (\ref{5}) and (\ref{1}),
see Table \ref{t1}.  As well as in case of
the {\em hfs} \cite{Terray}  the self energy
contribution is relatively large comparing to all other terms.
The last term ($(Z\alpha)^6m$), pure recoil contribution,  
has been recently calculated in Ref. \cite{Pachucki972}.
This correction is state dependent and amounts
\begin{eqnarray}
\Delta E(nS) = \frac{\alpha^6\,m}{n^3}\,&\biggl\{& -0.31056(63)
\nonumber \\ & & \hspace*{-5pt}
-\frac{1-n}{4\,n}+\frac{1-n^2}{3\,n^2}
-\frac{69}{512}\frac{1-n^3}{n^3}\biggr\}\,.
\end{eqnarray}
The sum of all constant $\alpha^6$ terms for 1S and 2S states is 
(see Table \ref{t1})
\begin{eqnarray}
E^{(6)}(1S) & = &  m\,\alpha^6\,\bigl[0.16104(64) - 
0.3868(22)\,\bbox{s}_-\bbox{s}_+\bigr]\,, \\
E^{(6)}(2S) & = &   \frac{m\,\alpha^6}{8}\,\bigl[0.34554(64) - 
0.0992(22)\,\bbox{s}_-\bbox{s}_+\bigr]\,.
\end{eqnarray}
The significant state dependence comes mainly from the 1--photon annihilation
contribution. As an example the $\alpha^6$ term  contributes 0.75 MHz
for $2^3S_1$ state.  

Unknown higher order terms limit the precision of theoretical predictions.
The double logarithmic term $\alpha^7\,\ln^2(\alpha)\,m$ is known
only for the hyperfine structure and annihilation contributions
\cite{Karshenboim931}, and is included in the theoretical value 
for the {\em hfs} in Table \ref{t2}. 
In the case of the spin independent part the double logarithmic 
contribution is unknown. We estimate it by 1 MHz for the ground
state by scaling the well known non-recoil correction to the hydrogen Lamb shift 
with the reduced mass factor $(\mu/m)^3$. 
 
The most sensitive on the current calculations is
the theoretical prediction for the  $1S-2S$ transition: 

\begin{equation} 
E(2^3S_1)-E(1^3S_1)=1233\,607\,221.0 (1.0)\; {\rm MHz}\,, \label{theory}
\end{equation}
where we use the values of the constant: 
$\alpha^{-1}=137.035\,999\,93(52)$ \cite{Kinoshita}
and $c\,R_\infty= 3\,289\,841\,960.394(27)$ MHz \cite{Udem}.
The result is in a moderate agreement with the experiment \cite{Fee}

\begin{equation}  
E(2^3S_1)-E(1^3S_1)=1233\,607\,216.4 (3.2)\; {\rm MHz}\,. \label{experiment}
\end{equation}
This new correction of the order  $\alpha(Z\alpha)^5\,m$
(Eqs. (\ref{4}) and (\ref{5})) contributes  significant amount of
-9.31 MHz to this transition. 
The comparison of theoretical predictions to a less precise 
$1S-2S$ measurement \cite{Danzmann} as well as for the $2S-2P$  experiments 
is presented in Table \ref{t2}. In all cases theoretical predictions are
more precise. The $2S-2P$ transitions
are weakly affected by the discrepancy in theoretical predictions,
{\em i.e.} by about 0.01 MHz. Experimental results of Refs. \cite{Hagena} and
\cite{Ley} are in agreement with theoretical predictions, while 
the results of Refs. \cite{Hatamian} and \cite{Mills75} are in a slight
disagreement. The hyperfine structure measurements are in agreement 
with the Caswell and Lepage result
\cite{Caswell78} and in 3.4 standard deviation disagreement 
with Ref. \cite{Pachucki971}. This disagreement could not be resolved
by comparison of the $1S-2S$ transition, since it is not sufficiently
sensitive on {\em hfs} contributions, see Eq. (\ref{8})

The evaluation of $(Z\alpha)^6\,m$ contribution was a complicated task.
Both calculations of positronium {\em hfs} at this order: 
Refs. \cite{Caswell78} and \cite{Pachucki971}
reproduce in the limit of heavy nucleus mass $M$
known results both in the orders of $(Z\alpha)^6m^2/M$ and
$(Z\alpha)^6m^3/M^2$ \cite{Bodwin}. 
Moreover, the result of Ref. \cite{Pachucki971}
is in agreement with the state dependent terms of these orders
calculated in Refs. \cite{Pyykko} and \cite{Sternheim} respectively.
The discrepancy between these  works will have to be resolved
before final conclusions could be drawn.

\bigskip

We wish to acknowledge Greg Adkins for  informing us on the 
results prior the publication and very fruitful discussions.
We thank  P. Labelle and J. Sapirstein for clarifying comments. 
Stimulating discussions with A. Yelkhovsky are
gratefully acknowledged by SGK.
We are grateful to  T. W. H\"ansch for the hospitality and support
during our stay at MPQ.  The work of KP was supported in part by the 
Polish Committee for Scientific Research under contract No. 2 P03B 024 11,
and the work of SGK was supported in part by the Russian State 
Program `Fundamental Metrology'.

\newpage


\begin{table}
\begin{tabular}{c|r|r|r|l}
&&&\\[-1ex]
Transition & Exp. Refs. & Experiment [MHz]& Theory [MHz] & Difference [MHz]\\[1ex]
\hline
\hline
&&&&\\[-1ex]
$2^3S_1-1^3S_1$ &\protect \cite{Danzmann} & 1233 607 218.9 (10.7) 
& 1233 607 221.0 (1.0) &  2.1(10.7)(1.0)\\[1ex]
&\protect \cite{Fee} & 1233 607 216.4 (3.2) & & 4.6(3.2)(1.0)\\[1ex]\hline
$1^3S_1-1^1S_0$ &\protect \cite{Mills83} & 203 387.5 (1.6)  
& 203 388.20 (0.80)  $^{a}$ &0.7(1.6)(0.8) $^{a}$\\[1ex]
                & & & 203 392.12 (0.50)  $^{b}$ & 4.6(1.6)(0.5) $^{b}$\\[1ex]
                &\protect \cite{Ritter} 
& 203 389.10 (0.74)& & -0.90(0.74)(0.80) $^{a}$\\[1ex]
                & &    &  & 3.02(0.74)(0.50) $^{b}$\\[1ex]\hline
$2^3S_1-2^3P_0$ &\protect \cite{Hatamian} & 18504.1 (10.0)(1.7) 
& 18498.42(0.13)  & -5.7(10.0)(1.7)\\[1ex]
                & \protect \cite{Hagena,Ley} & 18499.65 (1.20)(4.00)& 
& -1.2(1.2)(4.0)\\[1ex]\hline
$2^3S_1-2^3P_1$ & \protect \cite{Hatamian} & 13001.3 (3.9)(0.9) 
& 13012.58(0.13) & 11.3(3.9)(0.9)\\[1ex]
                &\protect \cite{Hagena,Ley} & 13012.42(0.67)(1.54) 
& & 0.2(0.7)(1.5)\\[1ex]\hline
$2^3S_1-2^3P_2$ &\protect \cite{Hatamian} & 8619.6(2.7)(0.9) 
& 8626.88(0.13) & 7.3(2.7)(0.9)\\[1ex]
&\protect \cite{Mills75} & 8618.4(2.8) & &8.5(2.8)\\[1ex]
&\protect \cite{Hagena,Ley} & 8624.38(0.54)(1.40) && 2.5(0.5)(1.4)\\[1ex]\hline
$2^3S_1-2^1P_1$ &\protect \cite{Conti} & 11181(13) & 11185.54(0.13) &5(13)\\[1ex]
&\protect \cite{Ley} & 11180(5)(4) &&6(5)(4)\\[1ex]\hline
$2^3S_1-2^1S_0$  & \multicolumn{2}{c|}{not measured yet} 
&25424.69(0.06) $^{b}$&\\[1ex]
\end{tabular}
\vspace{5mm}
\caption{
Comparision of experiments to theoretical predictions.
In case of the fine structure we give
statistical and systematic error separately because some data in Table are
correlated. Theoretical results for the ground state {\em hfs} 
were given in Refs. \protect \cite{Adkins971} and \protect \cite{Hoang}.
References to contradicting theoretical works are: 
$^a$ -- \protect \cite{Caswell86}; 
$^b$ -- \protect \cite{Pachucki971}. 
Theoretical uncertainties are dominated by estimates 
for unknown higher order terms.
The differences for $2S-2P$ are presented without theoretical uncertainties,
since they are significantly smaller than experimental ones. 
The levels are labeled as $n^{2S+1}L_J$.
       }
\label{t2}
\end{table}

\begin{table}
\begin{tabular}{c|c|c|c}
&&&\\[-1ex]
Contribution &  $K$& $\Delta E(1^1S_0)$ [MHz]&$\Delta E(1^3S_1)$ [MHz] \\[1ex]
\hline
\hline
&&&\\[-1ex]
$1\,\gamma$      & $-0.12565\,\left(\frac{3}{4}+\bbox{s}_-\bbox{s}_+\right)$   
& 0      & -2.34   \\[1ex]
$2\,\gamma$      & $ 0.03248\, \left(\frac{1}{4}-\bbox{s}_-\bbox{s}_+\right)$  
& 0.61   & 0       \\[1ex]
$3\,\gamma$      & $-0.05194\, \left(\frac{3}{4}+\bbox{s}_-\bbox{s}_+\right)$  
& 0      & -0.97   \\[1ex] 
$\alpha^2\,(Z\alpha)^4\,m$ & $0.02647-0.01374\,\bbox{s}_-\bbox{s}_+$         
& 0.69     & 0.43   \\[1ex]
$\alpha(Z\alpha)^5\,m$     & $0.5702(1)-0.5397(14)\,\bbox{s}_-\bbox{s}_+$    
& 18.19(2) & 8.12(1) \\[1ex]
$(Z\alpha)^6\,m$           & $-0.31056(63)+0.3767(17)\,\bbox{s}_-\bbox{s}_+$ 
&-11.07(2) & -4.04(1)\\[1ex]
\hline
&&&\\[-1ex]
total                      & $0.16104(64)-0.3868(22)\,\bbox{s}_-\bbox{s}_+$  
& 8.42(3)  & 1.20(2) \\[1ex]
\end{tabular}
\vspace{5mm}
\caption{All constant $\alpha^6\,m$ contributions to $1S$  state.  
The spin dependent part is similar to that in Ref. \protect\cite{Adkins971},
although we use a different naming.
Results are presented both, in relative units of 
$K\,\alpha^6\,m$ and in frequency units.}
\label{t1}
\end{table}


\begin{thebibliography}{99}
\bibitem{conventions} The natural units [$\hbar = c = 1$] are used
throughout. When $E$ is given in MHz it corresponds to transition frequency
$E/h$.
\bibitem{elsewhere} Details of this calculation will be presented elsewhere.
\end{thebibliography}
\end{document}